\documentclass[twocolumn,prl,aps,footinbib,amsmath,amssymb,superscriptaddress,longbibliography]{revtex4-1}
\usepackage{graphicx}
\usepackage{dcolumn}
\usepackage{xcolor}
\usepackage{bm}
\usepackage{natbib,hyperref}
\usepackage{amsmath,amssymb,amsfonts,bbold}
\usepackage[normalem]{ulem}
\usepackage{setspace}

 \definecolor{dartmouthgreen}{rgb}{0.05, 0.5, 0.06}
\definecolor{darkspringgreen}{rgb}{0.09, 0.45, 0.27}
\definecolor{DebianRed}{rgb}{0.84, 0.04, 0.33}
 \definecolor{darkpowderblue}{rgb}{0.0, 0.2, 0.6}

\newcommand{\beq}{\begin{equation}}
\newcommand{\eeq}{\end{equation}}
\newcommand{\beqa}{\begin{eqnarray}}
\newcommand{\eeqa}{\end{eqnarray}}

\begin{document}
\title{Anisotropic Ballistic Transport Revealed by Molecular Nanoprobe Experiments}
\author{Markus Leisegang}
\email{markus.leisegang@physik.uni-wuerzburg.de}
\affiliation{Physikalisches Institut, Experimentelle Physik II, Universit\"{a}t W\"{u}rzburg,
                Am Hubland, 97074 W\"{u}rzburg, Germany}
\author{Robert Schindhelm}
\affiliation{Physikalisches Institut, Experimentelle Physik II, Universit\"{a}t W\"{u}rzburg,
                Am Hubland, 97074 W\"{u}rzburg, Germany}
\author{Jens K{\"u}gel}
\affiliation{Physikalisches Institut, Experimentelle Physik II, Universit\"{a}t W\"{u}rzburg,
                Am Hubland, 97074 W\"{u}rzburg, Germany}
\author{Matthias Bode}
\affiliation{Physikalisches Institut, Experimentelle Physik II, Universit\"{a}t W\"{u}rzburg,
                Am Hubland, 97074 W\"{u}rzburg, Germany}
\affiliation{Wilhelm Conrad R{\"o}ntgen-Center for Complex Material Systems (RCCM),
                Universit\"{a}t W\"{u}rzburg, Am Hubland, D-97074 W\"{u}rzburg, Germany}

\begin{abstract}
Atomic-scale charge transport properties are not only of significant fundamental interest 
but also highly relevant for numerous technical applications.
However, experimental methods which are capable of detecting charge transport 
at the relevant single-digit nanometer length scales are scarce. 
Here we report on molecular nanoprobe (MONA) experiments on Pd(110)
where we utilize the charge carrier-driven switching of a single cis-2-butene molecule 
to detect ballistic transport properties over length scales of a few nanometers. 
Our data demonstrate a striking angular dependence with a dip in charge transport 
along the $\left[1 \bar{1} 0 \right]$-oriented atomic rows and a peak in the transverse $[001]$ direction.  
The narrow angular width of both features and distance-dependent measurements 
suggest that the nanometer-scale ballistic transport properties of metallic surfaces  
is significantly influenced by the atomic structure.   
\end{abstract} 
\date{\today}
\maketitle

%Introduction
{\em Introduction ---} The fate of elementary charges injected into a surface or interface 
is of fundamental interest for a myriad of technical applications.  
Due to their importance in CMOS transistors \cite{Sverdlov2008,Tagliabue2020,Duhan2020} 
and photovoltaics \cite{Gabor2011,Bakulin2012,Clavero2014,Burger2019}, 
ballistic hot electrons and holes, i.e., charge carriers with an energy 
well above the thermally broadened Fermi level, attracted particular attention. 
Immediately upon injection ballistic charge carriers occupy an intermediate state 
the properties of which are determined by the specific band structure of the receiving electrode \cite{Etheridge2019}.  
The dispersion relation of this band initially also dictates charge carrier propagation, 
until---after a few femtoseconds---a sequence of multiple weakly inelastic scattering events sets in, 
which drive it towards the band bottom and result in quasi-thermal equilibrium where diffusive transport dominates \cite{Etheridge2019}.  

Since building blocks of current electronics products often rely on functionalities 
where the atomic structure plays an important role, the investigation of ballistic transport on atomic length scales, 
where the interaction of charge carriers with single discontinuities could be investigated, would be highly beneficial. 
Conventional (single-tip) scanning tunneling microscopy (STM) and spectroscopy 
allow for the imaging of the surface structure and local density of states with atomic spatial resolution, respectively. 
However, ballistic transport properties can only be probed to a limited extent by the quasi-particle interference (QPI) technique 
because the tip simultaneously serves to inject and detect the charge carriers. 
As a result, QPI is restricted to transport pathways which exhibit a closed loop. 

To overcome this limitation several two-probe (2P) or even multiple-probe STMs 
have been designed, built, and utilized \cite{KHM2005,Miccoli2015,Yang2016,VCK2018}. 
A recent study demonstrated \cite{kolmer2019electronic} that the high stability of cryogenic setups
allows to position two tips of a 2P-STM on the same dimer row of a Ge(001)-c$(4 \times 2)$ surface.  
By driving a charge current through the intermediate sample, it became possible to observe 
one-dimensional ballistic transport down to an inter-probe distance $d_{\rm pp} = 30$\,nm \cite{kolmer2019electronic}.  
Even shorter distances are impeded by the spatial extent of the two probe tips, 
which---with good approximation---can be assumed as spherical with a typical diameter of 20 to 40\,nm \cite{Yang2016}. 

In an alternative approach, charge currents injected by the tip of a conventional, one-probe STM 
may be detected by molecular reactions \cite{maksymovych2007,sloan2010,ladenthin2015,schendel2016}.
Now the diameter of the only remaining STM tip is no longer a limiting factor. 
Reversible switching of a single molecule is utilized in the so-called molecular nanoprobe (MONA) technique, which allows for transport studies 
down to few-atom length scales \cite{kuegel2018imprinting,kugel2019impact,LKK2018,LBK2018}.
Furthermore, this technique allows to arbitrarily position the tip with respect to the detector molecule, 
whereas the shafts of the two probe tips in 2P-STM experiments may inhibit some configurations.  
This opens up the opportunity to investigate if and to what extent 
the atomic lattice of highly anisotropic surfaces impacts their transport properties.

In this study, we report on MONA experiments performed on the (110) surface of face-centered cubic (fcc) Pd. 
Utilizing a single cis-2-butene as a detector molecule we investigate the ballistic transport of charge carriers 
which are injected by an STM tip at probe--molecule distances of a few nanometers only. 
Our experimental data demonstrate a striking angular dependence. 
In particular, we observe a dip in charge transport along the $\left[1 \bar{1} 0 \right]$-oriented atomic rows 
and a peak in the transverse $[001]$ direction.  
The narrow angular width of both features and results obtained in distance-dependent measurements 
suggest that the nanometer-scale charge transport properties of metallic surfaces  
are significantly influenced by the atomic structure.   

{\em Experimental setup ---} Experiments were performed in a low-temperature LT-STM at a temperature $T \approx  4.5$\,K.   
The Pd(110) surface was prepared by cycles of Ar$^{+}$ sputtering 
with an ion energy of 0.7\,keV and subsequent annealing at 780\,K for 20\,min.
The cleanliness of the Pd(110) surface region for subsequent measurements is verified by an initial topographic scan 
before dosing minute amounts of cis-2-butene onto the surface inside the LT-STM at cryogenic temperature. 
% For topography images the LT-STM was operated in the constant-current mode 
% with the bias voltage $U$ applied to the sample.

\begin{figure}[t]   %%%%%%%%%%%%%%%%%%%%%%%%%%%%%%%%%%%%%%%%%%%%%%
	\includegraphics[width=\columnwidth]{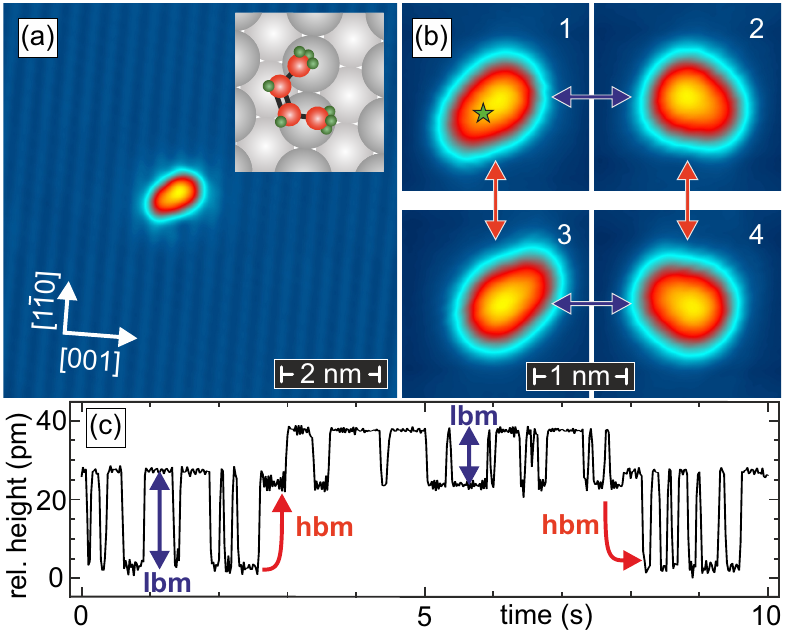}
		\caption{(a) Topographic STM image of a single cis-2-butene (c2b) molecule on the row-wise structure of a clean Pd(110) surface. 
		The inset schematically shows the molecule's adsorption geometry on top of a dense-packed Pd row. 
	        (b)~STM images of the four rotational states of c2b, labeled 1-4 (scan parameters: $U = 10$\,mV; $I = 20$\,pA). 
	        Blue arrows mark a low barrier motion (lbm) whereas red arrows mark a high barrie motion (hbm). 
	        (c) Telegraph noise recorded with the tip positioned at the green star in (b).  
	        Tunneling 	parameters: $U_{\rm bias} = 150$\,mV; $I = 20$\,pA.}
	 \label{Fig:Rotations}
\end{figure}

{\em Results ---} Figure\,\ref{Fig:Rotations}(a) shows constant-current STM image 
of a single cis-2-butene molecule (c2b) adsorbed on Pd(110). 
The substrate's atomic distance along the $[001]$ direction is by a factor of $\sqrt{2}$ larger 
than along the $\left[1 \bar{1} 0 \right]$ direction. 
This structural anisotropy of Pd(110) leads to a striped appearance in STM images, 
with adjacent $\left[1 \bar{1} 0 \right]$-oriented atomic rows being separated by 389\,pm. 
The inset in Fig.\,\ref{Fig:Rotations}(a) indicates the adsorption geometry of c2b 
on top of two neighboring atoms of a dense-packed Pd row \cite{sainoo2003observation}. 
When imaged by STM the c2b appears with an avocado shape, 
where the enlarged bright end depicts the two end-standing carbon atoms, 
whereas the elongated tale corresponds to the center carbon atoms inclosing the double bond. 

As first reported in Ref.\,\onlinecite{sainoo2005excitation} and confirmed by us, 
charge currents between the STM tip and the substrate through c2b can trigger two distinct transformations.
%the molecule can undergo two distinct transformations triggered by the application 
%of a specific threshold bias voltages with the tip positioned above c2b. 
At a bias voltage $U_{\rm lbm} \approx 30$\,mV a reversible rotation sets in 
which converts the molecule into its mirror image with respect to the $\left[1 \bar{1} 0 \right]$ axis, 
as marked by blue arrows in Fig.\,\ref{Fig:Rotations}(b). 
Owing to its low threshold energy it has been called low barrier motion (lbm) \cite{sainoo2005excitation}. 
At $U_{\rm hbm} \ge 100$\,mV another transition is observed which leads to a point inversion 
of the c2b molecule and will be denoted high barrier motion (hbm) hereafter.  
Within our measurement accuracy, the inversion point coincides 
with the point of maximum corrugation, probably representing the molecule binding site. 
%For the sake of clarity all four topographic scans in Figure\,\ref{Fig:Rotations}(b) are centered on the molecular maximum. 
Earlier experiments showed that both excitations are induced by inelastic electron tunneling \cite{sainoo2005excitation}. 
Detailed analysis revealed that the hbm is mainly driven by the C=C stretch mode, 
whereas the lbm is related to the Pd--C stretch mode.

   %%%%%%%%%%%%%%%%%%%%%%%%%%%%%%%%%%%%%%%%%%%%%%
These reversible transitions result in four molecular adsorption geometries, 
labeled as states 1 to 4 in Fig.\ \ref{Fig:Rotations}(b).
When choosing a suitable tip position where the four states are distinguishable, % in constant-current STM,  
such as the one marked by a green star in Fig.\,\ref{Fig:Rotations}(b), 
we can determine the actual state of the molecule by recording the tip height. 
An example of the resulting telegraph noise is shown in Fig.\,\ref{Fig:Rotations}(c). 
%The tunneling parameters were $U_{\rm bias} = 150$\,mV and $I = 20$\,pA. 
For $t \lesssim 2.6$\,s the apparent height fluctuates by about 25\,pm, indicating lbm between states 1/2. 
%This period is followed by a more rare hbm. %between levels 1/2 on one side and 3/4 on the other side.
Then, a more rare hbm to the high levels representing states 3/4 occurs, followed by lbm between these two states. 
Finally, at $t \approx 7.9$\,s another hbm sets the molecule back to states 1/2.
Our observations are on full accordance with previous findings reported in Ref.\,\cite{sainoo2005excitation}.

All data presented so far were measured with the STM tip positioned on top of the molecule. 
In order to investigate if the anisotropic structure of the Pd(110) surface 
also influences its charge transport properties in the ballistic regime
we conducted measurements with the MONA technique \cite{LKK2018}. 
In short, MONA utilizes a charge carrier-driven molecular switching event, 
such as a tautomerization or rotation, to detect currents injected a few nanometers away. 
Since the activation barrier of the molecular rotation essentially represents a high pass filter of the detection process,
only ballistic charge carrier, in this case with the threshold energy of the lbm, $eU_{\rm lbm} > 30$\,meV, can be detected. 
In this study we show that, by placing the STM tip and thereby the charge carrier injection point at different locations 
relative to the detector molecule,  the influence of an anisotropic atomic lattice on ballistic transport can be evaluated.
The measurement procedure is based on the cyclical repetition of three steps: 
First (i), the initial state of the molecule is probed by a topographic scan at non-invasive parameters ($U = 10$\,mV, $I = 20$\,pA). 
Then (ii), the tip is moved to the excitation position at a distance $r$ from the detector molecule 
and charge carriers are injected at excitation parameters $U_{\rm exc} = -50$\,mV and $I_{\rm exc} = 8$\,nA.  
Eventually (iii), the final state of the c2b molecule is probed by another topographic scan at non-invasive parameters. 
Switching events between the rotational states are detected by comparing the STM images before and after each injection. 
The switching probability is obtained by calculating the average of numerous measurements. 
To avoid the excitation of two or more switching processes with a single pulse, 
the probability per pulse is kept well below $15$\%.
The statistical standard variation of every data point presented in this work represents a minimum of 600 repetition cycles. 
Finally, we calculated the total electron yield $\eta$ by dividing the observed number of switching events $N_{\rm sw}$ 
through the amount of injected charge $N_{\rm el}$. 
The error bar for a calculated electron yield is given by $\delta n = \sqrt{\lbrack \eta (1-\eta) \rbrack / N_{\rm el}}$.

\begin{figure}    %%%%%%%%%%%%%%%%%%%%%%%%%%%%%%%%%%%%%%%%%%%%%%
	\includegraphics[width=\columnwidth]{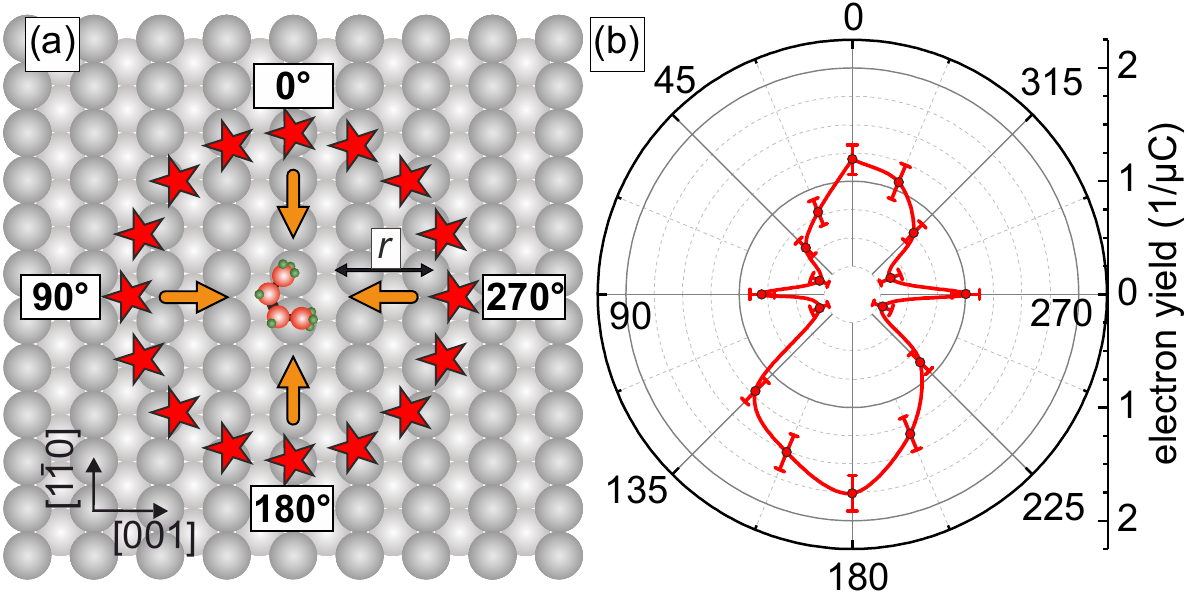}
		\caption{\label{MnPc_SbAg}
			(a) Schematic representation of the MONA measurements conducted 
			to detect the direction-dependent transport properties of the Pd(110) surface. 
			16 injection points, marked by red stars, are equidistantly arranged on a circle with radius $r = 3$\,nm 
			around a single cis-2-butene molecule, resulting in an angular resolution $\Delta \varphi = 22.5^{\circ}$. 
			Grey spheres represent the surface atomic structure of Pd(110) (not in scale).  
			(b) Polar plot showing the excitation angle-dependent electron yield 
			(excitation parameters: $U_{\rm exc} = -50$\,mV, $I_{\rm exc} = 8$\,nA, $t_{\rm exc} = 8$\,s). 
			The line serves as a guide for the eye only.}
		\label{Fig:Circle}	
\end{figure}      %%%%%%%%%%%%%%%%%%%%%%%%%%%%%%%%%%%%%%%%%%%%%%
In a first set of direction-dependent measurements, the switching rate of a single c2b molecule was investigated  
by injecting charge carriers at a constant distance of $r=3$\,nm from the molecule center under variation of the azimuthal angle.
The 16 excitation positions are represented as red stars in Fig.\,\ref{Fig:Circle}(a) 
which for comparison also shows the surface atomic structure of Pd(110) (not in scale).  
As indicated by orange arrows, measurements performed at $0^{\circ}$ and $180^{\circ}$ 
probe the transport properties along the $\left[1 \bar{1} 0 \right]$-oriented atomic rows of the Pd(110) surface, 
whereas data obtained at $90^{\circ}$ and $270^{\circ}$ give access to the transverse $\left[001 \right]$ direction.  
The results are shown in polar coordinates in Fig.\,\ref{Fig:Circle}(b). %, where the azimuthal angle defines 
%the direction of the tip relative to the detector molecule and the distance from the origin gives the total electron yield. 
On a qualitative level, we recognize a pronounced anisotropy with two main features: 
(I) Most apparent are broad maxima at $0^{\circ}$ and $180^{\circ}$, 
where the electron yield is four to six times higher than in the minima around $90^{\circ}$ and $270^{\circ}$. 
Furthermore, (II) the data exhibit more narrow maxima at $90^{\circ}$ and $270^{\circ}$ 
which are not as high as the values observed at $0^{\circ}$ and $180^{\circ}$, 
but still exceed the surrounding minima by roughly a factor of three 
\footnote{The atomic lattice presented in the background of Fig.\ \ref{Fig:Circle}(a) 
	suggests an electron yield which is mirror-symmetric with respect to the $\left[0 0 1 \right]$ 
	and the $\left[1 \bar{1} 0 \right]$ direction of Pd(110). 
	Whereas the data are mirror symmetric with respect to the $\left[1 \bar{1} 0 \right]$ direction,
	the broad lobe visible around $0^{\circ}$ is less extended than around $180^{\circ}$. 
	This asymmetry is caused by the fact that for the data presented in Fig.\,\ref{Fig:Circle}(b)
	exclusively switching events between states 1/2 connected by the lbm have been analyzed.
	As can be recognized in Fig.\,\ref{Fig:Rotations}(b), the elongated tale of the of these states
	point towards the bottom edge of the STM image, thereby breaking mirror symmetry with respect to the $\left[0 0 1 \right]$ axis.}. 

\begin{figure}[t]      %%%%%%%%%%%%%%%%%%%%%%%%%%%%%%%%%%%%%%%%%%%%%%
	\includegraphics[width=\columnwidth]{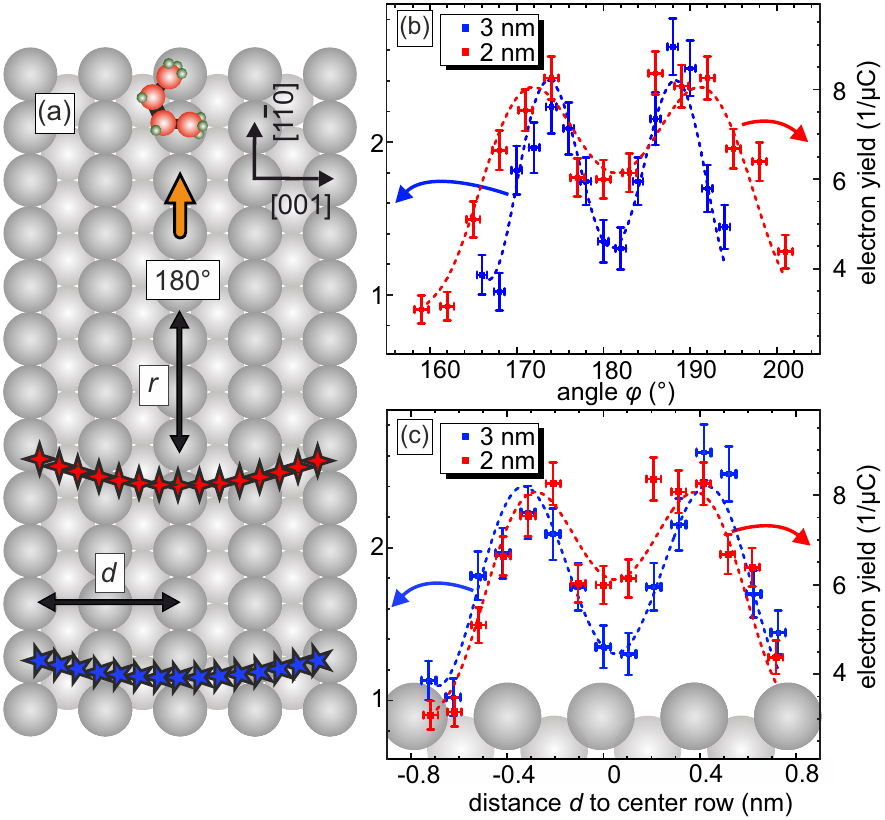}
		\caption{(a) Measurement scheme of transport measurements performed around the $\left[1 \bar{1} 0 \right]$ direction 
			at distances of $r = 2$\,nm (red) and $3$\,nm (blue) to the detector molecule.  
			%Each data set consist of 15 points with angular steps 
			%of $\Delta \varphi = 3^{\circ}$ and $\Delta \varphi = 2^{\circ}$ respectively.
			(b) Plot of the electron yield as a function of the angle $\theta$ relative to the $\left[1 \bar{1} 0 \right]$ direction.
			(c) Plot of the electron yield as a function of the distance $d$ of the injection point 
			from the adatom row where the c2b detector molecule is adsorbed 
			(excitation parameters: $U_{\rm exc} = -50$\,mV; $I_{\rm exc} = 8$\,nA, $t_{\rm exc} = 8$\,s). 
			A narrow minimum located at the central row can be recognized 
			which is surrounded by a symmetric double peak structure.}  
		\label{Fig:Onrow}		
\end{figure}

To shed light on the question whether the striking anisotropy observed in Fig.\,\ref{Fig:Circle} 
is driven by the dispersion relation or by scattering on single atomic rows, 
we performed MONA measurements at refined angle resolution. % around both high-symmetry exes.
The data obtained around the $\left[1 \bar{1} 0 \right]$ direction are presented in Fig.\,\ref{Fig:Onrow}. 
As sketched in Fig.\,\ref{Fig:Onrow}(a) we took two sets of data %where charge carriers were injected 
at tip--molecule distances of $r = 2$\,nm (red stars) and $r=3$\,nm (blue). 
In both cases 15 equidistant data points were taken, 
resulting in an angular resolution $\Delta \varphi = 3^{\circ}$ and $2^{\circ}$, respectively. 
Comparison with the surface lattice, which is represented as true-to-scale spheres in Fig.\,\ref{Fig:Onrow}(a), 
shows that the spacing of data points is well below the inter-row distance. 
%such a dense grid of data points shall allow for sub-atomic resolution across the dense-packed rows. 

      %%%%%%%%%%%%%%%%%%%%%%%%%%%%%%%%%%%%%%%%%%%%%%
Figure\,\ref{Fig:Onrow}(b) and (c) show the measured electron yield plotted versus 
the angle $\phi$ and the distance from the central dense-packed Pd row $d$, respectively. 
Based on the results obtained at low angular resolution in Fig.\,\ref{Fig:Circle}(b),
the findings presented here are quite surprising. 
Instead of a broad peak we now---at a much better resolution---recognize 
a narrow minimum located on the central row which is surrounded by a symmetric double peak structure.  
Comparing Fig.\,\ref{Fig:Onrow}(b) and (c) reveals that the data 
coincide significantly better in the representation over the distance $d$. 
Therefore, we conclude that the absolute charge carrier injection position on the row-wise structure 
rather than the relative direction of charge carrier propagation is relevant for the transport.

The observation of this narrow minimum suggests that the $\left[1 \bar{1} 0 \right]$-oriented adatom rows
posses a resistance which is enhanced as compared to the troughs in between.
Moving the charge carrier injection point away from this row through the intermediate trough 
to the next adatom rows leads to an enhanced transport as detected by an increasing electron yield. 
Only if the tip is moved further sideways beyond this adjacent adatom row 
a steep decline in the measured electron yield is detected.

\begin{figure}      %%%%%%%%%%%%%%%%%%%%%%%%%%%%%%%%%%%%%%%%%%%%%%
	\includegraphics[width=\columnwidth]{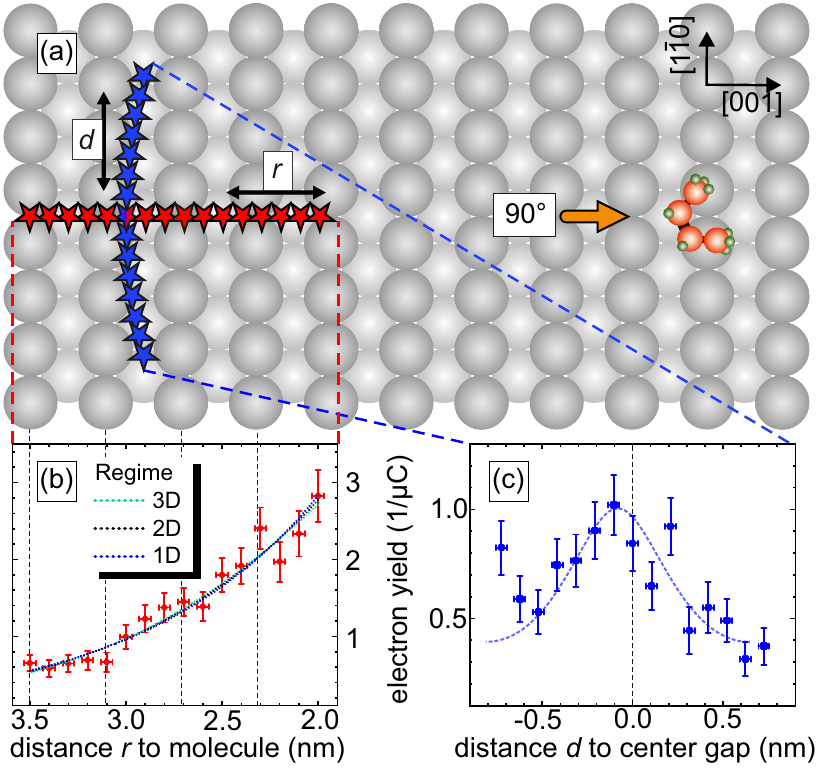}
		\caption{(a) Scheme of MONA measurements performed to determine 
			the distance dependence (red stars) and angular dependence around the $\left[0 0 1 \right]$ direction (blue).
			(b) Plot of the distance-dependent electron yield 
			for 2\,nm\,$\le r \le 3.5$\,nm with fits for 1D, 2D and 3D ballistic transport regime.
			(c) Electron yield as a function of the distance $d$ of the injection point 
			from the center gap row where the molecule c2b detector molecule is adsorbed.  
			Excitation parameters: $U_{\rm exc} = -50$\,mV; $I_{\rm exc} = 8$\,nA, $t_{\rm exc} = 8$\,s.}
	\label{Fig:Offrow}
\end{figure}      %%%%%%%%%%%%%%%%%%%%%%%%%%%%%%%%%%%%%%%%%%%%%%
In a second set of measurements with sub-atomic resolution we investigated the charge carrier transport 
along the $\left[0 0 1 \right]$ direction, i.e., transverse to the atomic rows.
As marked by red stars in Fig.\,\ref{Fig:Offrow}(a), we chose 16 equidistant excitation points 
at distances $r$ in the range between $r = 2$\,nm and $r = 3.5$\,nm from the molecular center. 
The respective electron yields are shown in Fig.\,\ref{Fig:Offrow}(b).
The distance $r$ dependence of the ballistic transport current $I_{\rm b}$ 
can be described by $I_{\rm b} = r^{-n} \cdot \exp\left( - r / L_{\rm inel} \right)$, 
where the first term depicts the intensity decay in $n+1$ dimensions and the second term describes 
the exponential damping due to inelastic scattering on the characteristic length scale $L_{\rm inel}$. 
Fitting the data to one- (1D), two- (2D), and three-dimensional (3D) transport 
results in very similar values of $0.5$\,nm\,$< L_{\rm inel}  < 1$\,nm, represented by dashed lines in Fig.\,\ref{Fig:Offrow}(b). 
This value is much lower than the inelastic mean free path usually found for very low electron energy, 
e.g., in noble metals \cite{Echenique2004}, and might be caused by enhanced $d$-band scattering in Pd \cite{ladstaeder2004}
which also inhibits the unambiguous identification of the dimension of the transport channel.
 
A complementary series of measurements was taken at 15 points around $90^{\circ}$ 
along a circular arc with a radius of $r=3$\,nm, see blue stars in Fig.\,\ref{Fig:Offrow}(a),
granting an angular resolution $\Delta \varphi = 2^{\circ}$. 
The resulting electron yield is displayed in Fig.\,\ref{Fig:Offrow}(c). 
We recognize a peak around the central atomic gap, $d = 0$\,nm, 
which quickly drops as the injection point is moved away from this position. 
Analysis reveals a peak width equivalent to the molecular adsorption site of about two atoms
%Therefore, we speculate that enhanced transport directly along the $\left[0 0 1 \right]$ direction 
%is possible without crossing atoms in the $\left[1 \bar{1} 0 \right]$ direction. 
\footnote{Again, the asymmetry around zero can be explained by the adsorption geometry of the molecule. 
	%At this low excitation bias only occasional hbm are observed, such that the analysis 
	Since the analysis is exclusively based on transitions between states 1/2, 
	the elongation of the c2b along the $\left[1 \bar{1} 0 \right]$ axis 
	breaks mirror symmetry with respect to the $\left[0 0 1 \right]$ axis.}.

{\em Discussion ---} 
The MONA data presented in Figs.\,\ref{Fig:Circle} to \ref{Fig:Offrow} unambiguously show  
that ballistic charge carrier transport on Pd(110) surfaces is strongly anisotropic. 
Measurements performed at relatively low angular resolution reveal that electron transport 
is more efficient along the $[1 \bar{1} 0]$ direction as compared to the $\left[0 0 1 \right]$ direction. 
This anisotropy might, in principle, be caused either by potential wells formed by the adatoms rows 
or by an anisotropic band structure of the Pd(110) surface or by a combination of both effects. 
High angular resolution scans reveal two very narrow features along the high symmetry directions, 
i.e., a double peak in the $[1 \bar{1} 0]$ direction and a single peak along $[001]$. 
The data presented in Figs.\,\ref{Fig:Onrow} and \ref{Fig:Offrow} evidence 
that the width of both features correlate much better to the real space atomic lattice 
than the angle under which the charge injection is performed.
These observations indicate that the adatoms, which create a row-wise structure on fcc(110) metal surfaces, 
serve as scattering potentials for the ballistic transport of hot charge carriers 
and that the band structure is of secondary relevance.

We speculate that the local density of states (LDOS) of the $d$-derived bands 
responsible for the unusual short IMFP of Pd \cite{ladstaeder2004} accumulates 
at the adatom rows along the $[1 \bar{1} 0]$ direction and decreases between them. 
This enhanced $d$-LDOS would lead to a potential well-like periodic variation of the electrostatic potential $V(r)$ 
with a high potential and strong scattering on top of the rows and potential troughs with much weaker scattering in between. 
Indeed, density functional theory calculations performed for Cu(110) also revealed a similar effect 
which results in a $V(r)$ modulation by a few hundred meV \cite{Stenlid2019}. 
As a result of this potential landscape, we measure a relatively low electron yield 
when the tip and the molecule are positioned on the same adatom row, 
since many electrons are scattered inelastically on their way from the injection point to the detector molecule.  
The assumption of scattering centers aligned along the adatom rows 
is also corroborated by the clear difference between the electron yield measure at $d = 0$\,nm 
for $r = 2$\,nm and $r = 3$\,nm, cf.\ Fig.\,\ref{Fig:Onrow}(c).  
For $r = 3$\,nm we recognize a deeper minimum, indicating that the larger number of scatterers 
present along the longer path result in a stronger reduction of the transport.  
Moving the tip away from the center row to the adjacent trough reduces scattering 
and results in an increased transport towards the molecule and, thereby, to an increased electron yield. 
Only if the injection point is moved across the next adatom row  
its additional scattering potentials become effective, resulting in a strong decline of the transport. 

Experiments along the $[001]$ direction further underpin the assumption of strongly localized scattering potentials. 
On the one hand we find that the inelastic mean free path (IMFP) of the charge carriers 
is on the order of one nanometer only, much shorter than what is usually observed for other noble metals. 
According to Ref.\,\onlinecite{ladstaeder2004}, the extraordinary reduction of the IMFP 
in Pd is caused by the strong contribution d-bands at the Fermi level. 
%Moreover, the calculations by \cor{Ladst{\"a}dter \it{et al}}.\ predict
%a reduced influence on high symmetry points where the d-band character is almost zero. 
%This could cause and additional enhancement of the anisotropy observed here.
On the other hand the data presented in Fig.\,\ref{Fig:Offrow}(c) reveal that scattering is reduced for charge carriers 
which propagate exactly perpendicular to dense-packed rows, as indicated by the higher electron yield. 
Therefore, we conclude that the row-wise atomic structure and the related surface potential
is the main factor which determines the transport properties of the Pd(110) surface. 
However, we cannot exclude that the anisotropic band structure also has some, though much weaker effect.

In summary, our results show that the charge carrier-induced switching 
of a single cis-2-butene molecule can be utilized to detect the ballistic transport properties 
of Pd(110) on length scales of a few nanometers.  
We find a rich angular dependence with two sharp features, i.e., 
a dip along the $\left[1 \bar{1} 0 \right]$-oriented atomic rows and a peak in the $[001]$ direction.  
These measurements provide evidence that the nanometer-scale charge transport properties 
of metallic surfaces is significantly influenced by scattering events on the atomic structure. 
We envision that similar measurements performed on materials more relevant for applications 
may not only lead to a better understanding of ballistic transport on atomic length scales 
but could also help optimizing the performance of nanoscale contacts and electrical junctions. 

\begin{acknowledgments}
We would like to thank S.\ Heinze and S.\ Haldar (Univ.\ Kiel) for insightful discussions.
We acknowledge financial support by the Deutsche Forschungsgemeinschaft (DFG, German Research Foundation) 
through grant BO 1468/27-1 and under Germany's Excellence Strategy through W{\"u}rzburg--Dresden Cluster of Excellence 
on Complexity and Topology in Quantum Matter ct.qmat (EXC 2147, project-id 390858490). 
\end{acknowledgments}

%\bibliography{anisotropic_transport_v7}

\end{document}